# Electron tunneling energies of a quantum dot in a magnetic field

S. Chaudhuri
Department of Chemistry and Physics
Monmouth University
400 Cedar Avenue
West Long Branch, NJ 07764
schaudhu@monmouth.edu

Theoretical analysis of the experimental energy data for an electron tunneling into a single electron state and to two interacting electrons state confined by a finite Gaussian potential in a 2D quantum dot subjected to a uniform magnetic field perpendicular to the plane of the dot is presented. While a previously published analytic solution for the magnetic field at which the 2-electron ground state transitions from the spin-singlet to the spin-triplet state agreed well with the data, the calculated energy for an electron tunneling into the two-electron state at higher magnetic fields diverged from the experimental data. The discrepancy was attributed to the magnetic field dependent Fermi energy of the $n^+$ electrode of the tunnel capacitor in the experiment. It is shown that the one electron and the two-electron experimental data agree remarkably well with the theory when we use a mechanism in which as if the added electron tunnels from a fixed energy state and the confining potential is a finite Gaussian potential in contrast to the generally used infinite harmonic confining potential. That the tunneling of an electron into the quantum dot from a fixed intermediate state assumed in the analysis is equivalent to the true tunneling picture where a mechanism known as Coulomb blockade that regulates the tunneling of one electron at a time into the dot and thus the measured electron tunneling energies are independent of the magnetic field dependent Fermi energy of the $n^+$ electrode is discussed.

## I.  INTRODUCTION

Wagner et al. [1] had predicted a transition for the ground state (g.s.) energy of a two-electron system in a two-dimensional quantum dot (2D QD) from the spin-singlet to the spin-triplet state as the magnetic field applied perpendicular to the plane of the dot increases. Ashoori et al. [2] experimentally observed such a transition. However, there was a significant discrepancy in the experimental value of the transition magnetic field with the theoretical value calculated by Wagner et al. Ashoori et al. conjectured that the discrepancy may be due to the assumption of strictly parabolic confining potential for the 2D QD used in the calculation. Recently Chaudhuri [3] presented an analytic solution to the problem with finite Gaussian confining potential for the QD. Indeed, the theoretical results for the finite confining potential model agrees well with the experimental value of the transition magnetic field. However, the electron addition energy calculated from the theoretical g.s. energy diverged from the experimental values at higher magnetic fields. Most of the theoretical treatments [4-15] have advantageously approximated the confining potential to be an infinite harmonic potential. Bruce and Maksym [8] treated the problem with a realistic confining potential including the effect of screening due to the gate electrodes using numerical diagonalization of the many-body Hamiltonian. While these calculations show the general nature of the dependence of the g.s. energy of n-electron systems on the magnetic field, none has provided a quantitative analysis of the experimental data of Ashoori et al. Using the analytic solution obtained in [3] we present a quantitative analysis of the 1-e and 2-e experimental electron tunneling data.

The electron addition energy measured by Ashoori et al. [2] was attributed to the electron energy corresponding to the lowest unoccupied electronic state of the dot being resonant with the Fermi energy of the n[+] electrode. The electrode also being subjected to the same magnetic field, the Fermi energy of the electrode should be dependent on the magnetic field. Therefore, it was considered that the theoretical electron tunneling energy calculation should incorporate the magnetic field dependent Fermi energy of the electrode in order to fit the experimental data with the theory. Chaudhuri [3] used an ansatz for the magnetic field dependent Fermi energy with Landau Level broadening and obtained a reasonable agreement with the experimental data at higher magnetic fields. The ansatz, however, did not include any term corresponding to de Haas-van Alphen (dHvA) oscillation. At the low temperature ($T = 0.35\ K$) at which the experiment was conducted one would expect the Fermi level to undergo some de Haas-van Alphen (dHvA) oscillation despite Landau Level (LL) broadening.

Recently, Chaudhuri [16] obtained an analytic solution for the Fermi energy with a Gaussian as well as a Lorentzian LL broadening and used it to fit the 2-e data. Even with the dHvA term dampened by the LL broadening, the fit still includes discernible oscillation that is not present in the experimental data. This indicates that the measured energies may not correspond with the energy required for direct tunneling from the n[+] electrode to the lowest unoccupied electronic state of the QD at the applied gate voltage. More importantly, however, the single electron energy data fit extremely well without any consideration of the magnetic field dependent Fermi energy and there is no indication of any oscillation in the single electron experimental data. If the tunneling of an electron into the two-electron state is from the highest occupied state of the electrode so must be the tunneling of an electron into the single electron state in the QD. However, the single electron energy data do not fit at all if we consider a direct tunneling between the magnetic field dependent Fermi level in the electrode and the 1-e g.s. in the QD. It clearly indicates that the experimental data do not depend on the magnetic field dependent Fermi energy of the n[+] electrode.

In this paper we present the theoretical results for the electron tunneling energies with a finite Gaussian QD confining potential. In the theoretical treatment we consider as if the added electron tunnels from a state of constant energy that does not dependent on the magnetic field. The theoretical results agree remarkably well with the experimental data. Ashoori [17] discussed in great detail the physics of the addition of an electron to a QD involving what is known as Coulomb blockade and the energetics involved in the process (see Ch. 6). It is shown that the tunneling of an electron from a fixed energy state into the QD considered in this paper is equivalent to the mechanism of Coulomb blockade regulating the tunneling of one electron at a time into the QD. Accordingly, the process used in the analysis is consistent with the true physical picture in which the electron addition energy does not depend on the Fermi energy of the n[+] electrode even though an electron transfers from the n[+] electrode to the QD when the Fermi energy of the QD is lowered by applying an external voltage below a certain equilibrium zone of the Fermi energy of the electrode [17].

## II. Experimental Data Analysis

We use Equations (16) and (17) in [3] to calculate the g.s. energies of a two-electron system and a single-electron system, respectively, in a 2D QD confined by a Gaussian potential $V(r_i) = V_0\left(1 - e^{-\alpha^2 r_i^2}\right)$, where $r_i$ is the *i-th* electron coordinate. The corresponding harmonic potential is given by $V_h = \frac{1}{2}m^*\omega_0^2 r_i^2$, *where* $\omega_0$ is defined by letting $\alpha \to 0$ while holding $V_0 \alpha^2 = \frac{1}{2}m^*\omega_0^2$ constant, where $m^*$ is the effective electron mass. See reference [3] for details. We obtained the experimental 1-e and 2-e data by digitizing the two lowest curves of Fig. 2 in the paper by Ashoori et al. [2]. The energy scale of the experimental data is set by the vertical bar representing an energy of 5 meV. We should note that the experimental 1-e energy value at zero magnetic field is indefinite. We assumed an arbitrary energy for the 1-e data at zero magnetic field which is later fixed by setting it to the calculated 1-e g.s. energy. During the digitization we had set the bottom of the 5 meV energy bar to be at 0. Consequently, the values of the experimental 1-e and 2-e data at zero magnetic field were 4.1 meV and 8.5 meV, respectively.

We denote the digitized values of the 1-e and 2-e experimental data by $y_1(B)$ *and* $y_2(B)$, respectively, and the theoretical g.s. energies of the 1-e and 2-e systems calculated by using Equations (17) and (16) in [3] by $e_{g1}(B)$ *and* $e_{g2}(B)$, respectively. We set the value of $y_1(0)$ at the calculated 1-e g.s. energy $e_{g1}(0)$ to obtain the normalized experimental data given by the equation

$$y_{n1}(B) = y_1(B) + e_{g1}(0) - y_1(0). \tag{1}$$

Consequently, the normalized 2-e experimental data is given by

$$y_{n2}(B) = y_2(B) + e_{g1}(0) - y_1(0). \tag{2}$$

The difference $y_2(B) - y_1(B)$ represents the amount of energy arising from Coulomb interaction between the two electrons in the 2-e state. In order to establish the relationship of $y_{n2}(B)$ with the calculated 2-e g.s. energy, $e_{g2}(B)$, we note that the second electron tunnels into the 2-e state of the QD in which there is already an electron in the 1-e g.s. from an intermediate state with a fixed energy $e_r$. The energy corresponding to the gate voltage at which the second electron tunnels into the QD 2-e state from the intermediate state with the fixed energy $e_r$ is the normalized 2-e experimental value. Thus, the theoretical energy $e_{t2}(B)$ corresponding to the experimental normalized energy, $y_{n2}$, is given by

$$e_{t2}(B) = e_{g2}(B) - e_{g1}(B) - e_r. \tag{3}$$

Assuming that $e_r$ is constant i.e., the energy of the intermediate state from which the electron tunnels into the QD does not depend on the magnetic field, we obtain $e_r$ by equating $y_{n2}(0)$ given by Eq. (2) with the added energy given by Eq. (3) *at* $B = 0$ as

$$e_r = [e_{g2}(0) - 2e_{g1}(0)] - [y_2(0) - y_1(0)]. \tag{4}$$

We fit the normalized experimental data $y_{n2}(B)$ with the calculated theoretical values of $e_{t2}(B)$ by varying the two Gaussian potential parameters. Instead of using $V_0$ and $\alpha$ for the Gaussian potential we use the parameters $\varepsilon_0$ and $\alpha_0$ defined by $\varepsilon_0 = \sqrt{2V_0}\alpha_0$ and $\alpha_0 = \frac{\hbar}{\sqrt{m^*}}\alpha$. Then we used the same parameters that fit the 2-e data to calculate the theoretical 1-e g.s. energy $e_{g1}(B)$ using Eq. (17) in [3].

In Fig. 1 we show the experimental and the corresponding theoretical results for the electron tunneling energy to the 1-e and the 2-e states. The circles in the upper curve represent the normalized experimental data $y_{n2}(B)$ and the solid line represents the fitted theoretical data $e_{t2}(B)$. The values of the parameters for the least square fit are $\varepsilon_0 = 10.96\ meV$ and $\alpha_0 = 2.43\ \sqrt{meV}$. Note that $e_r$ is not a fit parameter. The value of $e_r$ according to Eq. (4) is determined by the theoretical 1-e and 2-e g.s. energies calculated with the potential obtained by the best fit parameters $\varepsilon_0\ and\ \alpha_0$ and the difference between the 2-e and 1-e experimental data at $B = 0$. The calculated value of $e_r$ is 6.1 meV. The singlet-to-triplet transition magnetic field is at 1.97 T. The circles in the lower curve represent the normalized 1-e experimental data $y_{n1}(B)$ and the solid line represents the calculated theoretical values of $e_{g1}(B)$ using the values of $\varepsilon_0\ and\ \alpha_0$ obtained from the 2-e data fit above. Since the lower curve represents the 1-e g.s. energy $e_{g1}(B)$, the zero energy is set at the bottom of the QD confining potential. Both 1-e and 2-e data fit extremely well with a single set of the parameter values $\varepsilon_0$ and $\alpha_0$ defining the Gaussian QD confinement potential. According to this hypothesis the experimental data for the n-electron system ($n \geq 2$) should correspond to $e_{tn}(B) = e_{gn}(B) - e_{g(n-1)}(B) - e_r$.

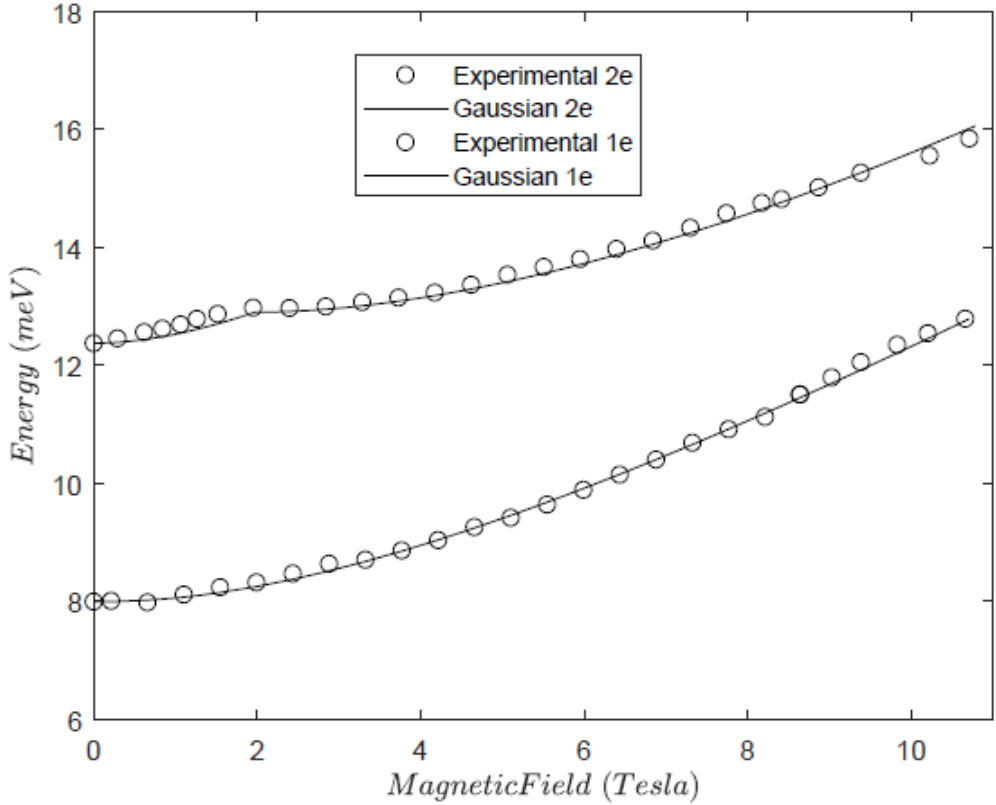

FIG. 1. Experimental and theoretical energies vs magnetic field for the 1-e and 2-e systems. The circles in the upper curve represent the normalized experimental data $y_{n2}(B)$ and the solid line represents the fitted theoretical data $e_{t2}(B)$ for the Gaussian confining potential. The values of the parameters for the least square fit are $\varepsilon_0 = 10.96\ meV$ and $\alpha_0 = 2.43\ \sqrt{meV}$. The circles in the lower curve represent the normalized 1-e experimental data $y_{n1}(B)$ and the solid line represents the calculated theoretical values of $e_{g1}(B)$ for the same Gaussian potential.

In Fig. 2 we show the various energy values including the energy value $e_r$ obtained from the data fit. The 1-e and 2-e g.s. energies, $e_{g1}(0)$ and $e_{g2}(0)$, at zero magnetic field are 8.0 meV and 26.5 meV, respectively. The energies to add the first electron and the second electron corresponding to the experimental data, $y_{n1}(0)$ and $y_{n2}(0)$ are 8.0 meV and 12.4 meV, respectively. The value of $e_r$ is 6.1 meV. All energies are with respect to the bottom of the confining potential of the QD.

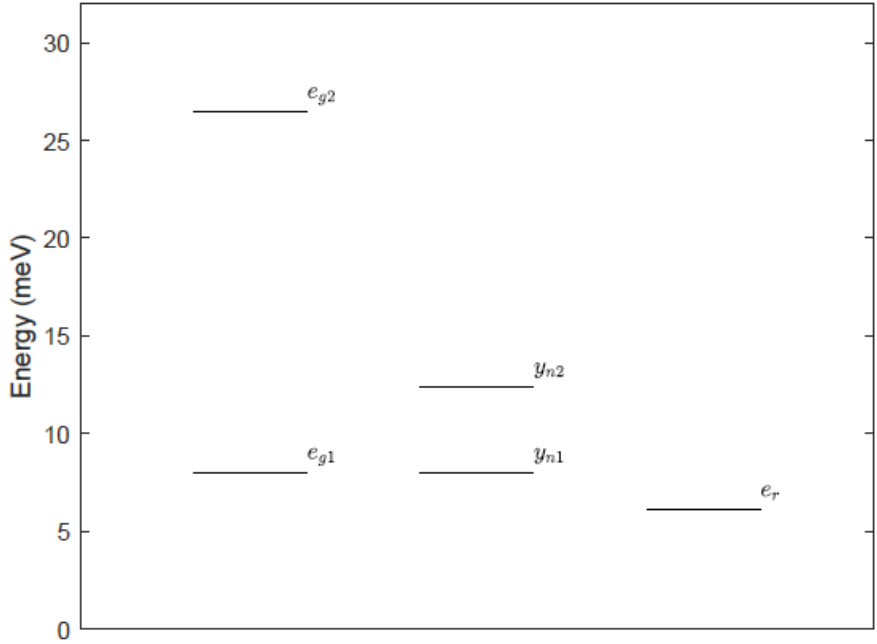

FIG. 2. Energy values of the 1-e and 2-e systems at zero magnetic field. The theoretical g.s. energies $e_{g1}(0) = 8.0\ meV$ and $e_{g2}(0) = 26.5\ meV$ of the 1-e and 2-e electron systems are shown in the leftmost column. The energies to add the first electron and the second electron corresponding to the experimental data, $y_{n1}(0) = 8.0\ meV$ and $y_{n2}(0) = 12.4\ meV$, are shown in the second column. The energy of the intermediate state from which the electrons tunnel into the QD, $e_r = 6.1\ meV$, is shown in the third column.

We attempted to fit the experimental data with the assumption that an electron tunnels directly from the n[+] electrode. We used an analytic solution for the LL broadened Fermi energy [16] of the n[+] electrode. We find that there is still some dHvA oscillation at the broadening parameter used to fit the 2-e data. However, there is no discernible dHvA oscillation in the experimental data that we would expect despite the LL broadening at the low temperature of 0.35 °K. More importantly, the 1-e data do not fit at all if we assume a magnetic field dependent state from which the electron directly tunnels into the QD 1-e state. The excellent fit of both 1-e and 2-e energies with a single set of parameters and no indication of a magnetic field dependent state from which the first electron tunnels into the 1-e state or the second electron tunnels into the 2-e state strongly support the tunneling mechanism in which as if the added electron tunnels from a constant energy state.

We used the constant energy, $e_r$, in our analysis of the experimental data as if it is the energy of an intermediate state from which an electron tunnels into the 2-e state of the QD for a simpler visualization of the process used in the analysis above. In fact, it is related to the mechanism known as Coulomb blockade regulating the tunneling of one electron at a time into the QD. Ashoori [17] described in great detail (see Ch. 6) the physics and the energetics of adding an electron to the dot by applying an external gate voltage. According to this mechanism, by applying an external voltage the Fermi energy of the dot must be lowered than that of the electrode by an amount $e^2/C + \delta_n$, where $\delta_n$ is the difference in energy between the lowest unoccupied state and the highest occupied state of the QD for the addition of the *n-th* electron, where $C$ is the capacitance of the dot to its surroundings, and the energy $e^2/C$ is a manifestation of the so-called Coulomb blockade. Once the electron tunnels into the dot, the Fermi energy of the dot rises by the amount $e^2/C + \delta_n$. The 1-e and 2-e experimental energy data [2] correspond to the gate voltages at which the first and the second electron, respectively, tunnels into the dot. With $\delta_2 = e_{g2}(B) - e_{g1}(B)$, the energy $(e^2/C + \delta_n)$ corresponds to $e_{t2}(B)$ given in Eq. (3) and $e_r$ is related to $e^2/C$ within an additive constant. The Fermi energy of the electrode changes with the applied magnetic field. However, the applied voltage lowers the Fermi energy of the dot below the "equilibrium zone" with respect to the electrode Fermi energy at the given magnetic field to induce the dot to take another electron. Thus, the electron tunneling energy $e^2/C + \delta_n$ measured by the gate voltage does not depend on the electrode Fermi energy as if the electron tunnels from a state with the fixed energy, $e_r$ with respect to the QD potential bottom regardless of the Fermi energy of the electrode.

In Fig. 3 we show the QD confining potential. The solid line represents the Gaussian potential $V(\mathbf{r})$ with the fitted parameter values $\varepsilon_0 = 10.96 \; meV \; and \; \alpha_0 = 2.43 \; \sqrt{meV}$. The dashed line represents the corresponding harmonic potential $V_h$. The dotted and the dash-dotted lines represent the values of $\hbar\omega_0$ corresponding to the harmonic potential, i.e., $\alpha \to 0$, and $V_0$, respectively.

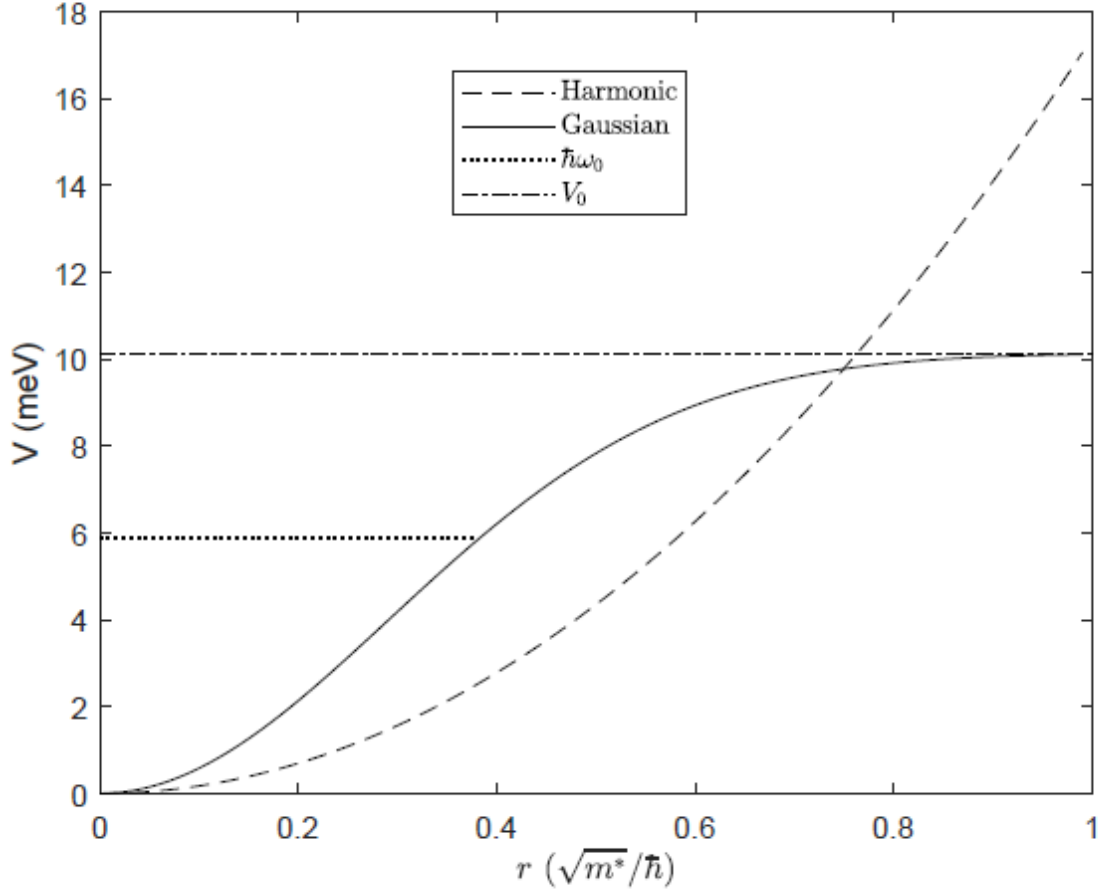

FIG. 3. The QD confining potential. The solid line represents the Gaussian potential $V(r)$ with the fitted parameters $\varepsilon_0 = 10.96\ meV$ and $\alpha_0 = 2.43\ \sqrt{meV}$. The dashed line represents the corresponding harmonic potential $V_h$. The dotted and the dash-dotted lines represent the values of $\hbar\omega_0 = 5.86\ meV$ corresponding to the harmonic potential, i.e., $\alpha \to 0$, and $V_0 = 10.1\ meV$, respectively.

In Fig. 4 we show the experimental and the corresponding theoretical data with the harmonic potential for the 1-e and the 2-e systems. The circles in the upper curve represent the normalized experimental data $y_{n2}(B)$ and the solid line represents the fitted theoretical data $e_{t2}(B)$. The harmonic potential parameter $\hbar\omega_0 = 5.86\ meV$ for the least square fit. The circles in the lower curve represent the normalized 1-e experimental data $y_{n1}(B)$ and the solid line represents the calculated theoretical values of $e_{g1}(B)$. The energy of the intermediate state from which the electrons tunnel into the QD, $e_r$, is 4.1 meV. The 1-e data fit is still excellent while the 2-e data at higher magnetic field is worse than that with the Gaussian confining potential. The 1-e state wave function being confined closer to the center while the wave function for the 2-e state being spread out from the center of the QD due to the Coulomb interaction, the effect of finite confining potential is less for the 1-e state compared to the 2-e state. Consequently, we expect the impact of the finite potential to the 2-e state to be more pronounced than to the 1-e state. The magnetic field for the singlet-to-triplet state with the Gaussian and the harmonic potentials are 1.97 and 2.72 T, respectively. We should note that the 2-e g.s. singlet-to-triplet transition magnetic field did not change from the value obtained in [3] despite the tunneling mechanism used in this paper is different. The reason, of course, is that the energy difference between the singlet and the triplet states of the 2-e g.s. does not depend on the tunneling mechanism.

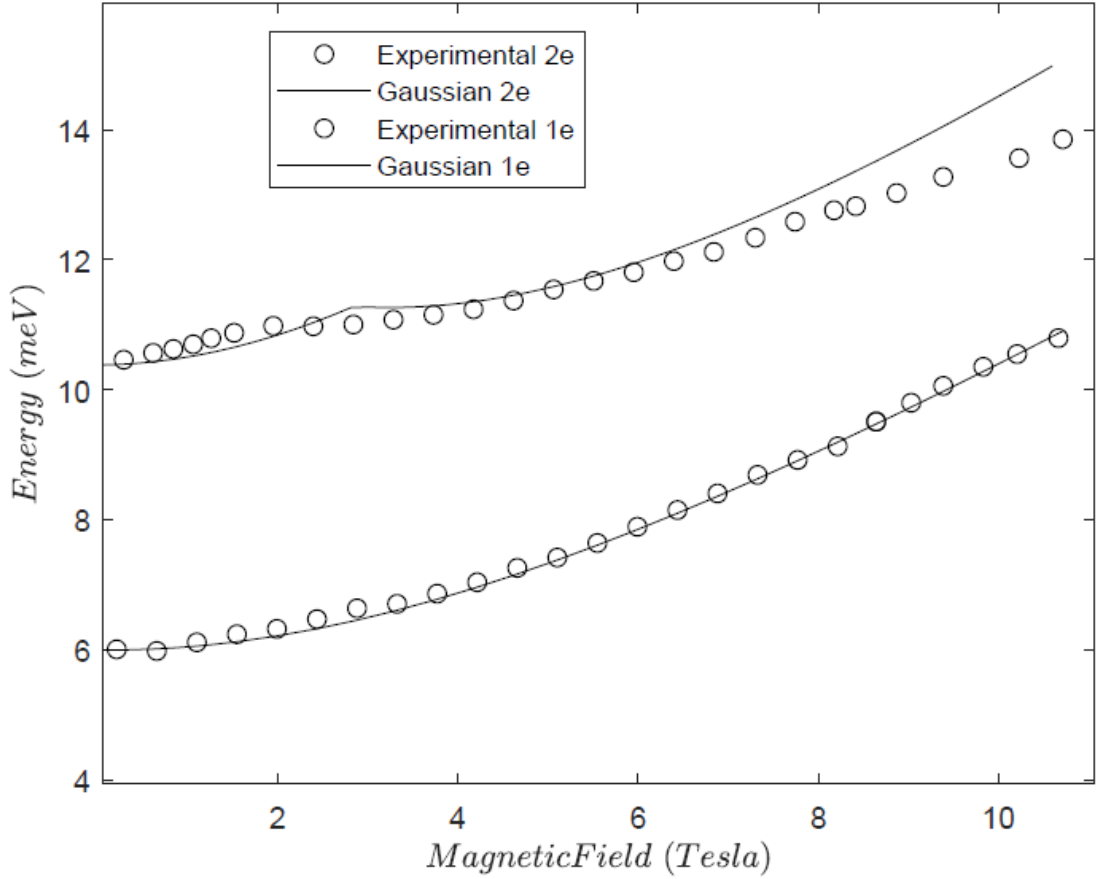

FIG. 4 The 1-e and the 2-e g.s. energies versus magnetic field for the harmonic confining potential. The circles in the upper curve represent the normalized experimental data $y_{n2}(B)$ and the solid line represents the fitted theoretical data $e_{t2}(B)$. The harmonic potential parameter $\hbar\omega_0 = 5.86\ meV$ for the least square fit. The circles in the lower curve represent the normalized 1-e experimental data $y_{n1}(B)$ and the solid line represents the calculated theoretical values of $e_{g1}(B)$ obtained with $\hbar\omega_0 = 5.86\ meV$.

### III. Conclusion

Using an analytic result recently obtained for the LL broadened Fermi energy of a 3D electron gas as a function of the applied magnetic field we found that the experimental data do not support the current hypothesis that electron tunneling occurs directly between the $n^+$ electrode and the 2D QD and that the electron tunneling energy should depend on the electrode Fermi energy. Instead, the data fit exceedingly well with the theory in which a more realistic finite Gaussian QD confining potential compared to infinite harmonic potential and an electron tunneling mechanism from a fixed energy intermediate state are used. The tunneling mechanism used in the analysis is equivalent to the true physical mechanism known as the Coulomb blockade that regulates an electron tunneling into the QD. Both the 1-e and 2-e experimental data fit with a single set of two fit parameters representing the Gaussian confining potential. Also is presented the data fit using a harmonic confining potential for the QD. The effect of the finite confining potential compared to the infinite harmonic potential is clearly demonstrated by a significantly better agreement of the 2-e g.s. singlet-to-triplet transition magnetic field as well as of the 2-e electron tunneling energy data at higher magnetic field values with the experimental data.

## ACKNOWLEDGMENTS

It is a pleasure to thank R. Ashoori at the Massachusetts Institute of Technology for providing a copy of his Ph.D. thesis.